\documentstyle[preprint,tighten,aps,amstex,graphicx,times,floats,epsfig]{revtex}
\def\ba{\begin{eqnarray}}
\def\ea{\end{eqnarray}}
\def\bq{\begin{equation}}
\def\eq{\end{equation}}
\def\lsim{\mathrel{\raisebox{-.6ex}{$\stackrel{\textstyle<}{\sim}$}}}

\begin{document}

\thispagestyle{empty}

\newcommand{\sla}[1]{/\!\!\!#1}

\preprint{
\font\fortssbx=cmssbx10 scaled \magstep2
\hbox to \hsize{
\hskip.5in \raise.1in\hbox{\fortssbx University of Wisconsin - Madison}
\hfill\vtop{\hbox{\bf MADPH-00-1191}
            \hbox{\bf IFT--P.076/2000}
            \hbox{\bf September 2000}
            } 
}
}

\vskip.5in

\title{ Observing an invisible Higgs boson }

\author{
O.~J.~P.~\'Eboli$^1$ and D.~Zeppenfeld$^{2}$
} 

\address{ 
$^1$Instituto de F\'{\i}sica Te\'orica -- UNESP \\ 
             R. Pamplona 145, 01405-900 S\~ao Paulo, Brazil\\
$^2$Department
 of Physics, University of Wisconsin, Madison, WI 53706, USA
}  

\maketitle 

\begin{abstract}

   Given its weak coupling to bottom quarks and tau leptons, the Higgs boson
   may predominantly decay into invisible particles like gravitinos,
   neutralinos, or gravitons. We consider the manifestation of such an
   invisibly decaying Higgs boson in weak boson fusion at the CERN
   LHC. Distinctive kinematic distributions of the two quark jets of the
   signal as compared to $Zjj$ and $Wjj$ backgrounds allow to restrict the
   Higgs branching ratio to 'invisible' final states to some 13\%
   with 10~fb$^{-1}$ of data, provided events with two energetic forward 
   jets of high dijet invariant mass and with substantial missing 
   transverse momentum can be
   triggered efficiently.  It is also possible to discover these particles
   with masses up to 480~GeV in weak boson fusion, at the 5$\sigma$ level,
   provided their invisible branching ratio is close to 100\%.

\end{abstract} 


\newpage

\section{Introduction}

Some extensions of the Standard Model (SM) exhibit Higgs bosons which can
decay into stable neutral weakly interacting particles, therefore giving rise
to invisible final states. In supersymmetric models the Higgs bosons can decay
with a large branching ratio into lightest neutralinos or gravitinos in some
region of the parameter space \cite{inv1}, leading to an invisible final state
if $R$ parity is conserved. Invisible Higgs decays also happen in models with
an enlarged symmetry breaking sector, {\em e.g.} in Majoron models
\cite{inv2,inv3}, where the Higgs disintegrates into light weakly interacting
scalars. In the scenario of large extra dimensions, proposed by Arkani-Hamed,
Dimopoulos and Dvali \cite{adv}, it is possible that the Higgs boson mixes
with scalar fields arising from gravity propagating in extra dimensions
\cite{inv4}. This mixing can lead to a sizeable invisible width for the Higgs.

Certainly the presence of invisible decays modifies considerably the Higgs
boson searches, making it much more difficult. At $e^+e^-$ colliders the
problem is not very severe, and it has been shown that the Higgs parameter
space can be probed completely \cite{epem}. Presently, the LEP II
collaborations exclude invisible Higgs masses up to 106.7 GeV
\cite{exp}. On the other hand, the Higgs search at hadron colliders is much
more difficult in the presence of such invisible decays. Previous studies have
analyzed $ZH$ and $WH$ associated production\cite{h:zh} and $t \bar{t} H$
production\cite{h:tt} as promising channels. Considering statistical errors
only and assuming that the Higgs couplings are the SM ones and that the
invisible branching ratio effectively is 100\%, the associated $ZH$ production
was estimated to be sensitive to Higgs masses $m_H \lsim 150$ GeV \cite{h:zh}
at the CERN LHC while $t \bar{t} H$ production might extend the Higgs mass
range to 250 GeV \cite{h:tt}. Both searches must deal with signal cross
sections in the few fb range or below, however, and detector resolution and
background normalization errors become crucial because signal and background
distributions have virtually identical shapes after cuts.

In this work we show that the LHC's potential to unravel the existence of
invisibly decaying Higgs bosons can be considerably extended by studying Higgs
production via weak boson fusion. Weak boson fusion (WBF) is a copious source
of Higgs bosons at the LHC, yielding signal cross sections, after cuts, of
order 100 fb, as we shall see.  Furthermore, the presence of very energetic
forward jets in the signal, with characteristic azimuthal angle correlations,
allows us to tag the Higgs events and to efficiently suppress the
backgrounds. It has been shown that WBF is among the most promising search
channels in the 120--200 GeV Higgs mass range. It is possible to observe an
intermediate mass Higgs boson through its decays $H \to \gamma\gamma$
\cite{dd:aa}, $H \to W^* W^* \to e^\pm \mu^\mp \sla{p}_T$ \cite{dd:ww}, and 
even $H \to \tau^+ \tau^-$ \cite{dd:tt}. Here we extend this list to include
otherwise invisible decay modes.

This paper is organized as follows. In Section II, we describe the techniques
used to evaluate the relevant cross sections, while Section III contains the
main characteristics of the signal and backgrounds and the cuts chosen to
enhance the invisible Higgs signal. In Section~IV, we present a detailed
discussion of how to use data to predict the exact background normalization.
Our results and conclusions are presented in Section V.

\section{Calculational tools}

We are considering the production of Higgs bosons in WBF, $qq\to qqVV \to qqH$
($V=W$ or $Z$), with subsequent decay to undetectable particles. The signal is
thus characterized by two quark jets, which typically enter in the forward
and backward regions of the detector and are widely separated in
pseudorapidity, and by a large transverse momentum imbalance, due to the 
Higgs' invisible decay products. Significant backgrounds can arise from any 
processes leading to two jets and missing transverse momentum, such as
$Zjj$ production with subsequent decay $Z\to \bar\nu\nu$ or $Wjj$ production
with $W\to\ell\nu$ decay where the charged lepton is not identified. 
Another source are QCD
dijet and multijet events with large missing transverse momentum generated by
energy mismeasurements or from high $p_T$ particles escaping through the
beam-hole. These purely QCD processes possess a potentially very large cross 
section. In mismeasured two jet events, the missing $p_T$
points in the direction of the jets and we use this fact to suppress this
background.  However, this is not true for the production of 3 or more jets
and we include this QCD $jjj$ background in our analysis.

The signal and backgrounds are simulated at the parton level with full tree
level matrix elements.  This was accomplished by numerically evaluating
helicity amplitudes for all subprocesses. Backgrounds include all order
$\alpha_s^2$ real emission corrections to Drell-Yan production (to be called
QCD $Wjj$ and $Zjj$ processes) and cross sections are calculated with code
based on Ref.~\cite{BHOZ}.  The second large class of processes are the signal
and the electroweak background, EW $Wjj$ and $Zjj$ production, with
contributions from the $t$-channel exchange of an electroweak boson off which
the final state $W$, $Z$ or Higgs is radiated. The code for these processes is
based on Ref.~\cite{CZ}, and was checked with matrix elements generated using
the package Madgraph \cite{mad}. Madgraph code was also used to simulate the
QCD $jjj$ background at tree level. For all QCD effects, the running of the
strong coupling constant is evaluated at one-loop order, with $\alpha_s(M_Z) =
0.12$. We employed CTEQ4L parton distribution functions \cite{CTEQ4_pdf}
throughout. The factorization scale was chosen as $\mu_f =$ min($p_T$) of the
defined jets. We took the electroweak parameters $\sin^2 \theta_W = 0.23124$,
$\alpha_{em} = 1/128.93$, $m_Z = 91.189$ GeV, and $m_W = 79.95$ GeV, which was
obtained imposing the tree level relation $\cos
\theta_W = m_W/m_Z$. For the QCD $jjj$ background, for which detector effects
produce the missing $p_T$, we simulate experimental resolutions by smearing
the energies (but not directions) of all final state partons with a Gaussian 
error given by $\Delta(E)/E = 0.5/\sqrt{E} \oplus 0.02$ ($E$ in GeV).

$W\to l\nu$ decays in $Wjj$ events lead to a $jj\sla p_T$ signature when the
charged lepton $l=e,\mu,\tau$ is not identified. A precise modeling of this
background requires a full detector simulation. We estimate the $Wjj$
backgrounds by assuming that in the central region, $|\eta_l|<2.5$, all muons,
electrons and taus with $p_T(l)>5,10,20$~GeV, respectively, can be vetoed,
while any charged leptons below these thresholds will be misidentified and
counted in the $p_T$ balance only.  In the forward regions, $2.5<|\eta_l|<5$,
a lepton veto is taken to be impossible. Here, muons are assumed to give no
$p_T$ deposit in the calorimeters, in contrast to electrons and taus whose
entire energy is recorded. Note that the resulting $Wjj$ background, within
jet cuts given below, is about
half the event rate of all $Wjj, W\to l\nu$ events with $p_T(\nu)>100$~GeV,
i.e. we are certain not to seriously underestimate the $Wjj$ background.

An important feature of the WBF signal is the absence of color exchange
between the final state quarks, which leads to a depletion of gluon emission
in the region between the two tagging jets. We can enhance the signal to
background ratio by vetoing additional soft jet activity in the central region
\cite{veto}. A central jet veto is ineffective against the EW $Wjj$ and $Zjj$
backgrounds which possess the same color structure as the signal.  For the QCD
backgrounds, however, there is color exchange in the $t$-channel and
consequently a more abundant production of soft jets, with $p_T>20$~GeV, in
the central region \cite{CZ}. The probability of an event to survive such a
central jet veto has been analyzed for various processes in
Ref.~\cite{rainth}, from which we take the veto survival probabilities of
Table~\ref{surv} which are appropriate for the hard tagging jet cuts to be
used below.

The cross section for Higgs boson production via WBF is well known within the
framework of the SM. We should keep in mind that this production cross section
might be diluted in extensions of the SM. For instance, it is suppressed by
factors $\sin^2(\beta-\alpha)$ or $\cos^2(\beta-\alpha)$ in supersymmetric
models. Any suppression in the production cross section has the same effect,
for our study, 
as a branching ratio of invisible Higgs decays below unity, and we 
will not separate these effects in the following.

\section{Signal and  Background Properties}

The main features of the production of an invisible Higgs boson via WBF are 
the presence of two very energetic forward jets as well as a large missing 
transverse momentum.  Therefore, we initially impose the following jet 
tagging cuts and missing momentum cut
\begin{eqnarray}
&& p_T^j > 40 \hbox{ GeV} \;\;\;\;\;\;\; , \;\;\;\; | \eta_j | < 5.0 
\nonumber
\\
&& | \eta_{j1} - \eta_{j2} | > 4.4 \;\;\;\; , \;\;\;\;   
\eta_{j1} \cdot \eta_{j2} < 0 \; ,
\label{cuts1}
\\
&& \sla{p}_T > 100 \hbox{ GeV} \; .
\label{cuts1a}
\end{eqnarray}
A further reduction of the backgrounds, with good signal efficiency, is 
achieved by requiring a large invariant mass, $M_{jj}$, of the two
tagging jets, 
\begin{equation}
M_{jj} > 1200 \hbox{ GeV} \; ,
\label{cuts2}
\end{equation}
and by selecting events where the azimuthal angle between the tagging jets,
$\phi_{jj}$ (measured in radians) is relatively small, 
\begin{equation}
\phi_{jj} < 1   \; ,
\label{cuts3}
\end{equation}

\begin{figure}
\protect
\centerline{\mbox{\epsfig{file=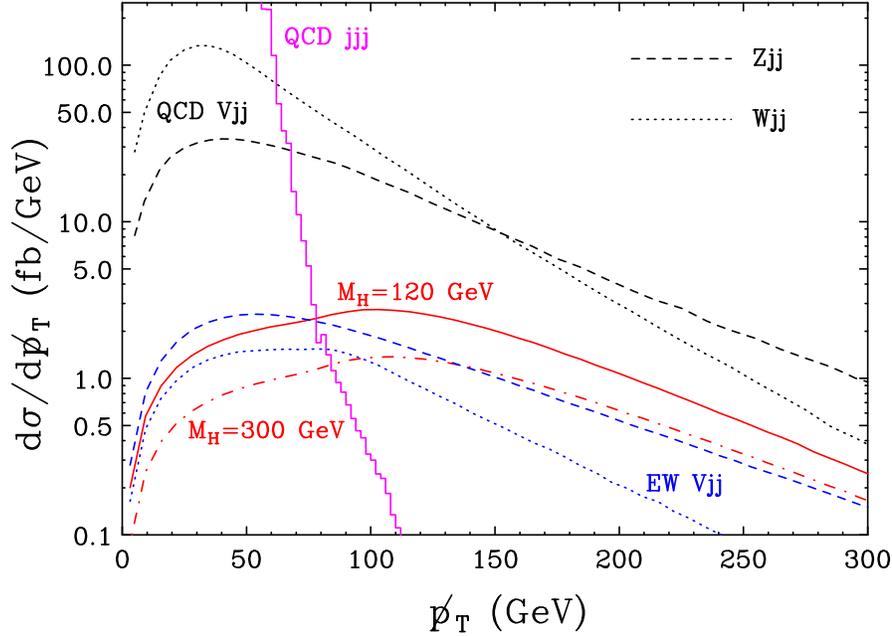,angle=90,width=0.75\textwidth}}}
\caption{
Missing transverse momentum spectra within the cuts (\protect\ref{cuts1}) and
(\protect\ref{cuts2}).
Results are shown separately for the EW $Zjj$ (blue dashed line) and 
$Wjj$ (blue dotted line) backgrounds, as well as the QCD processes 
$Zjj$ (black dashed line), $Wjj$ (black dotted line), and $jjj$ 
(magenta histogram) production. We exhibit the invisible Higgs 
contribution for $M_H=120$ (red solid line) and 300 GeV 
(red dot-dashed line).  
}
\label{dist:et}
\end{figure}

In order to motivate our choice of the $\sla{p}_T$ cut, we display, in Fig.\ 
\ref{dist:et}, the $\sla{p}_T$ spectrum after the cuts (\ref{cuts1})
and (\ref{cuts2}), 
but without a central jet veto. The signal
exhibits a peak around $\sla{p}_T \simeq 100$ GeV and it is much smaller than
the backgrounds at small $\sla{p}_T$. Missing $p_T$ generated by the 
QCD $jjj$ background falls rapidly and this background becomes negligible 
above $\sla p_T=100$~GeV. Note that we require $\phi_{jj}<2.6$ for the two 
tagging jets of the QCD jjj background, in order to avoid the soft 
singularities present near $\phi_{jj}=\pi$. Well above 
$\sla{p}_T \simeq 100$~GeV, 
the missing $p_T$ spectra of the signal and the $Zjj$ backgrounds 
have the same slope. Hence, a tightening of the $\sla p_T$ cut soon becomes
useless.

\begin{figure}
\protect
\centerline{\mbox{\epsfig{file=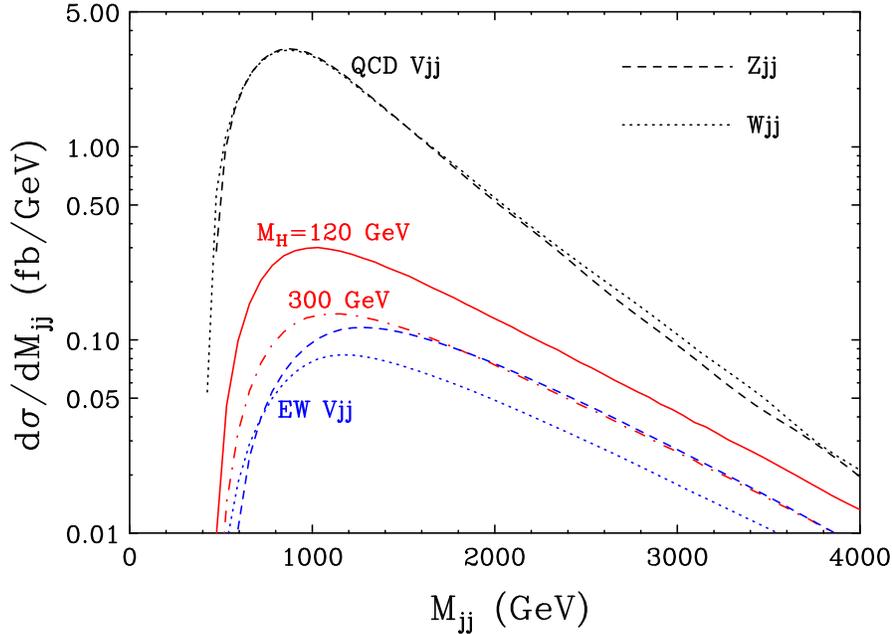,angle=90,width=0.75\textwidth}}}
\caption{
  Dijet invariant mass distributions when applying the cuts of 
  Eqs.~(\protect\ref{cuts1},\protect\ref{cuts1a}).
  The lines follow the same convention as in Fig.~\protect\ref{dist:et}.
}
\label{dist:mjj}
\end{figure}

The QCD backgrounds involve initial and final state gluons which tend to be
softer than the quarks in WBF. In Fig.~\ref{dist:mjj}, this is reflected by 
the steeper falloff of the QCD backgrounds as $M_{jj}$, the dijet invariant 
mass, is increased. The $M_{jj} > 1200$ GeV requirement
reduces these backgrounds sufficiently. Note that no central jet veto is 
included in Fig.~\ref{dist:mjj}. A further improvement of the
signal to background ratio is possible by tightening the $M_{jj}$ cut, but 
this will not be pursued in the following.

\begin{figure}
\protect
\centerline{\mbox{\epsfig{file=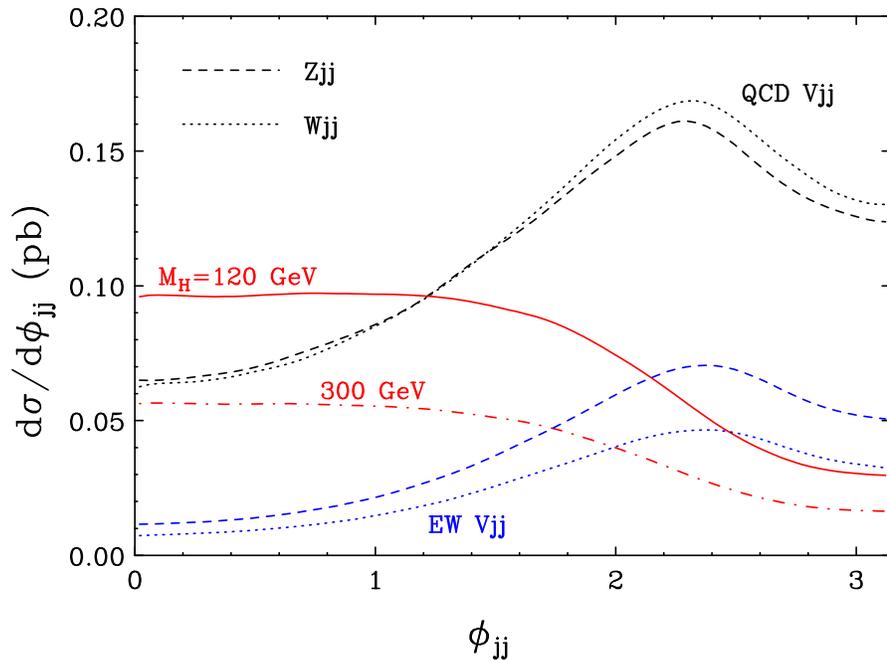,angle=90,width=0.75\textwidth}}}
\caption{
Distributions of the azimuthal angle separation between the two tagging 
jets for the various background processes and the Higgs signal at
$M_H=120$ and 300~GeV. Results are shown after applying the cuts
(\protect\ref{cuts1}-\protect\ref{cuts2}) and including the effect of a 
central jet veto with the survival probabilities of Table~\protect\ref{surv}.
The lines follow the same convention as in Fig.~\protect\ref{dist:et}.  
}
\label{dist:phijj}
\end{figure}

The most distinct remaining difference between the Higgs signal and all 
backgrounds is the azimuthal angle correlation of the two tagging jets.
The $\phi_{jj}$ distributions within the cuts (\ref{cuts1}-\ref{cuts2})
and including the central jet veto efficiencies of Table~\ref{surv}
are shown in Fig.~\ref{dist:phijj}. The $hV_\mu V^\mu$ coupling of the Higgs
boson in WBF favors Higgs emission opposite to both tagging jets, which
leads to small values of $\phi_{jj}$. The $Wjj$ and $Zjj$ backgrounds, on 
the other hand, are smallest in this region and prefer back to back jets. 
The distinct shape difference
of the $\phi_{jj}$ distributions provides very powerful tools to probe
for a Higgs contribution, either by a full shape analysis or by the 
$\phi_{jj}<1$ cut of (\ref{cuts3}).

The effect of such a cut is presented in Tables \ref{ev:bckg} and 
\ref{ev:sig}, where background and signal cross sections are given after 
imposing the cuts (\ref{cuts1}-\ref{cuts2}) and (\ref{cuts1}-\ref{cuts3}).
With an integrated luminosity of 10 fb$^{-1}$ a total of 1670 background and
of order 400 to 1000 signal events with $\phi_{jj}<1$ are expected for an
invisible branching fraction Br$(H\to {\rm invisible})=1$, giving a highly
significant signal when statistical errors only are considered.

\section{Predicting the background}

Finding, or constraining, an invisibly decaying Higgs boson signal in
$jj\sla p_T$  events is essentially a counting experiment since a resonance 
in the invariant mass distribution of the Higgs decay products cannot be 
extracted. The sensitivity of the search is thus determined by the precision
with which the background rate in the search region can 
be predicted. Since the signal selection is demanding, including double 
forward jet tagging and central jet vetoing techniques whose acceptance 
cannot be calculated with sufficient precision in perturbative QCD, the 
background levels need to be determined directly from LHC data.

Fortunately, a sizable sample of $Vjj$ events ($V=W$ or $Z$), with
fully identified charged leptons from the $V$ decay, will be available within
the hadronic acceptance cuts discussed above. For $Zjj$ events with two 
identified charged leptons, the $\sla p_T>100$~GeV cut is equivalent to 
$p_T(Z)>100$~GeV and the only difference to the $Zjj,\;Z\to\nu\bar\nu$ 
background is due to minimal charged lepton $p_T$ and rapidity cuts which 
are needed to insure their observability. The cross section for
QCD and EW $Zjj,\;Z\to\l^+l^-$ events ($l=e,\;\mu$), with $p_T(l)>15$~GeV,
$|\eta_l|<2.5$ and within the cuts (\ref{cuts1}-\ref{cuts2}), including the
central jet veto probabilities of Table~\ref{surv}, is about 87 fb, of which
16.6~fb are expected in the $\phi_{jj}<1$ region. Thus an effective luminosity
of 10~fb$^{-1}$ is sufficient to measure the $Zjj$ background in the two
regions $\phi_{jj}>1$ and $\phi_{jj}<1$ with relative errors of 3.8\% and
7.8\%, respectively. 

The level of the $Wjj$ background can be extracted from $W^\pm\to l^\pm\nu$
events, with hadronic cuts as given above and the additional requirements
$p_T(W)>100$~GeV and $p_T(l)>25$~GeV, $|\eta_l|<2.5$. The combined cross
section for QCD and EW $(W\to l\nu)jj$ events, including central jet veto,
is 184~fb (975~fb) with (without) the $\phi_{jj}<1$ requirement. With an 
effective luminosity of 10~fb$^{-1}$ this translates into a statistical error
of a mere 2.3\% for the prediction of the total $Wjj$ background within
the cuts (\ref{cuts1}-\ref{cuts3}). Central
jet veto efficiencies will be almost identical for charged lepton and missing
$p_T$ signatures of the decaying $W$s and $Z$s. Other systematic errors like
luminosity uncertainties, jet reconstruction efficiencies, or
knowledge of the fractional
contributions from WBF and QCD production are also eliminated by
obtaining the background normalization from the leptonic $W,\;Z$ data.
Even trigger efficiencies can be determined directly from these events.

The background normalization error can be further reduced by expanding the
calibration region, at the price, however, of introducing a QCD uncertainty,
due to the necessary extrapolation to the signal region. For the $Zjj$ 
background in particular, one would like to use the entire $\phi_{jj}$
region to determine the background normalization, since the data sample within
(\ref{cuts3}) will be relatively small. These extrapolation
uncertainties are small, as we will now show for the shape of the $\phi_{jj}$
distribution.

\begin{figure}[thb]
\begin{center}
\includegraphics[width=15.0cm,angle=90]{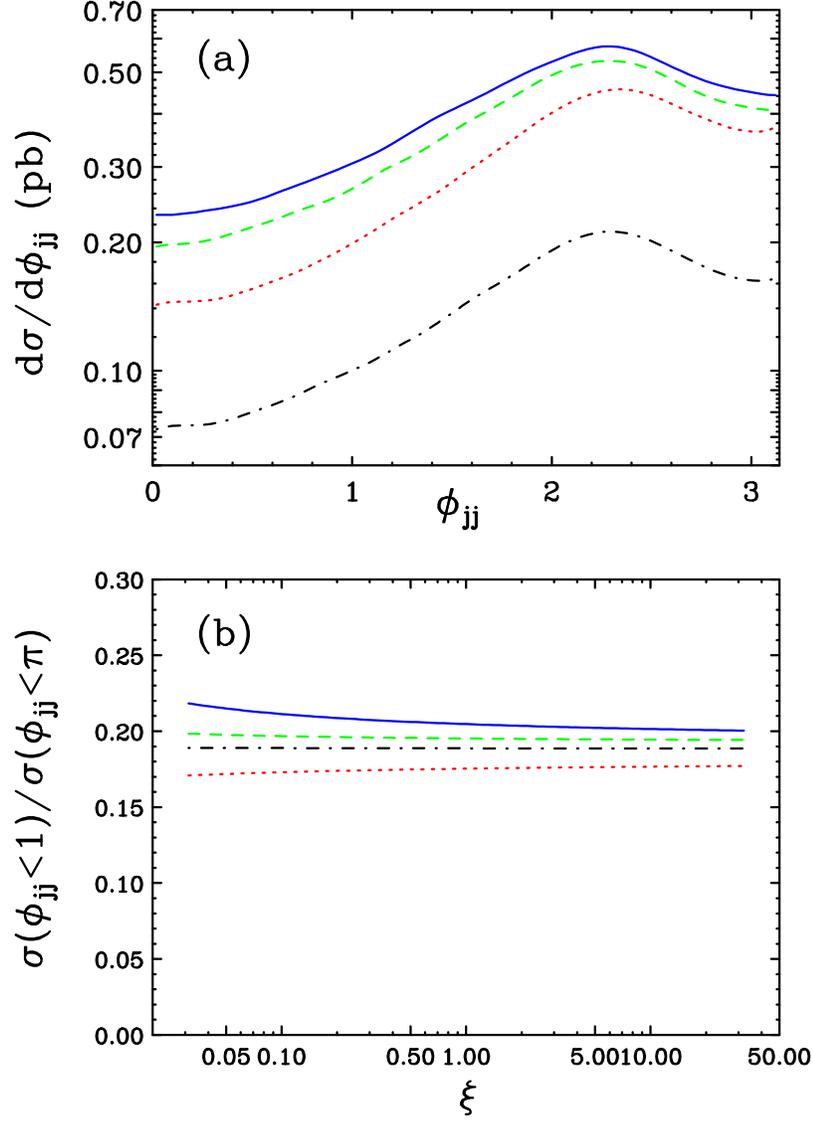}
\end{center}
\vspace*{0.2cm}
\caption{Scale dependence of the shape of the dijet azimuthal angle 
distribution of QCD $Zjj$ events with $Z\to\nu\bar\nu$ in LO QCD. 
Results are shown for four choices of the basic 
renormalization scale, $\mu_R^0=\sqrt{\hat s/4}$
(dash-dotted line), $\mu_R^0=\sqrt{(E_T^2(Z)+p_{Tj_1}^2+p_{Tj_2}^2)/3}$ 
(green dashed line), the default choice
$\alpha_s^2(\mu_R^0) = \alpha_s(p_{Tj_1})\;\alpha_s(p_{Tj_2})$
(blue solid line), and $\mu_R^0=E_T(Z)$ (red dotted curve). In (b) the
fraction of events with $\phi_{jj}<1$ is shown as a function of $\xi$,
where $\mu_R=\xi\mu_R^0$.}
\label{fig:scale}
\vspace*{0.2cm}
\end{figure}

At present we only have leading order (LO) calculations of the $Wjj$ and $Zjj$
QCD backgrounds available. Due to the small difference in weak boson mass, as
compared to {\em e.g.} the large dijet mass required in our event selection,
QCD corrections for these processes are expected to be very similar and we
only analyze the shape of the $\phi_{jj}$ distribution for the QCD $Zjj$
background in the following. Shown in Fig.~\ref{fig:scale}(a) is
$d\sigma/d\phi_{jj}$ for four different renormalization scale choices,
$\mu_R^0=\sqrt{\hat s/4}$ (dash-dotted line) where $\hat s$ is the squared
parton center--of--mass energy,
$\mu_R^0=\sqrt{(E_T^2(Z)+p_{Tj_1}^2+p_{Tj_2}^2)/3}$ (dashed line), the default
choice $\alpha_s^2(\mu_R^0) = \alpha_s(p_{Tj_1})\alpha_s(p_{Tj_2})$ (solid
line), and $\mu_R^0=E_T(Z)$ (dotted curve). One finds that the normalization
of the QCD background changes by up to a factor of 3 between these choices and
another variation by a factor of 3 to 4 is obtained by changing individual
renormalization scales between $\mu_R = \mu_R^0/10$ and $\mu_R =
10\mu_R^0$. While the normalization of the QCD $Zjj$ cross section changes
drastically, the shape of the $\phi_{jj}$ distribution is essentially
unaffected. As a measure of shape changes we plot the 
fraction of events with $\phi_{jj}<1$,
\begin{equation}
R_1 = {\int_0^1 {d\sigma\over d\phi_{jj}}\;d\phi_{jj} \over
\int_0^\pi {d\sigma\over d\phi_{jj}}\;d\phi_{jj}}\; ,
\label{eq:R1}
\end{equation}
as a function of $\xi$, the scale factor for the four different renormalization
scale choices $\mu_R = \xi\mu_R^0$ listed above. The $\xi$ dependence shown
in Fig.~\ref{fig:scale}(b) is very small for individual choices of $\mu_R^0$,
smaller in fact, than the differences between the four basic scales $\mu_R^0$.

Fig.~\ref{fig:scale}(b) indicates that $R_1 = 0.19 \pm 0.02$ at LO QCD, i.e. a
scale uncertainty of at most 10\% is found for the shape of the $\phi_{jj}$
distribution, a very small QCD error indeed for a leading order
calculation. While this error is still uncomfortably large for the actual
determination of an invisible Higgs contribution to $jj\sla p_T$ events, a NLO
calculation, which hopefully will be available by the time the experiment is
performed, should push the QCD shape uncertainty to well below the 5\%
level. We assume a 5\% QCD uncertainty on $R_1$ in the following.

Compared to this extrapolation from opposite side to same side
dijet events, minimal changes
in hadronic event properties are encountered when extrapolating from $Vjj$
events with observed leptons, as discussed above, to the corresponding 
$jj\sla p_T$ sample. 
Hence QCD uncertainties for ratios like $\sigma(Zjj,Z\to\nu\bar\nu)/
\sigma(Zjj,Z\to\l^+l^-)$, within analog cuts, 
are expected to be small. Indeed, an analysis of the scale 
uncertainties of these ratios, analogous to the one performed above for $R_1$,
points to LO QCD errors of $3\%$ or less for the $Wjj$ ratios and 1-2\%
for the $Zjj$ ratios. At NLO these QCD uncertainties should be entirely
negligible and we do not consider them in the following.

\section{Discussion}

The previous discussion points to at least two methods for constraining, or
discovering, an invisible Higgs decay channel. The first uses the shape of the
$\phi_{jj}$ distribution only. With 10~fb$^{-1}$ already, the estimated 1670
background events in the $\phi_{jj}<1$ region imply that $R_1$ in
(\ref{eq:R1}) can be measured with a statistical error of 2.2\%, which is
already small compared to the systematic error of 5\% due to QCD scale
uncertainties. Taking the expectations for an $M_H=120$~GeV Higgs boson as an
example (see Table~III), an invisible effective branching fraction
$B_H=\hbox{Br}(H\to {\rm invisible})$ would give rise to a modification
\begin{equation}
R_1 = {1670+967B_H\over 9180+2380B_H} = 0.182(1+0.32B_H+\dots)\;.
\end{equation}
In absence of a signal, a deviation of $R_1$ of more than 10.7\% can be ruled
out at 95\% CL, which translates into an upper bound
\begin{equation}
B_H < 34\% \qquad \hbox {at 95\% CL}\;,
\end{equation}
from shape information alone. Clearly this bound is systematics dominated
and can be improved if it can be shown that our estimate for the QCD scale
uncertainty is too conservative.

A much higher sensitivity is achieved by making use of charged lepton
signatures for $Vjj$ events, as discussed in the previous section. Again
assuming an integrated luminosity of 10~fb$^{-1}$, the $Wjj$ background can
directly be determined from $W\to l\nu$ events in the region $\phi_{jj}<1$,
with a statistical error of 2.3\%. The number of $Zjj$ events with 
$Z\to l^+l^-$, in this region, is still modest, which makes it 
worthwhile to combine the direct determination with an extrapolation 
of the $(Z\to l^+l^-)jj$ cross section in the $\phi_{jj}>1$ region, in 
spite of the 5\% systematic uncertainty of $R_1$. Combining all errors 
in quadrature, the
$(Z\to\nu\bar\nu)jj$ background in $\phi_{jj}<1$ can be predicted, directly
from the data, with a 5.3\% error. The combined background from $Wjj$ and
$Zjj$ sources can thus be predicted with an accuracy of $3.0\%$.  Combined
with the expected statistical uncertainty of the 1670 background events, one
finds that the Higgs contribution can be measured with a $1\sigma$ error of
$6.4$~fb, which, in absence of a signal, translates into a 95\% CL bound of
\begin{equation}
B\sigma(qq\to qqH,\;H\to{\rm invisible}) < 12.5\;{\rm fb}.
\label{limit}
\end{equation}
For an integrated luminosity of 100 fb$^{-1}$ the combined backgrounds can be
obtained with a precision of 1.2\% and the bound (\ref{limit}) reduces to 4.8
fb. We present in Table~\ref{res} the invisible Higgs branching ratio that can
be probed at 95\% CL as a function of $M_H$. In the event that the invisible
Higgs production cross section is suppressed with respect to the SM, the
limits in this table apply to the product of the invisible branching ratio
times the suppression factor. We find that the WBF channel extends considerably
the sensitivity for invisibly decaying Higgs bosons, with respect to the $ZH$
and $\bar{t} t H$ channels \cite{h:zh,h:tt}. With 10~fb$^{-1}$
(100~fb$^{-1}$) of data it is possible to discover these particles with masses
up to 480~GeV (770~GeV), at the 5$\sigma$ level, provided their invisible
branching ratio is 1.

A disadvantage of the WBF process, as compared to $ZH$ and $\bar{t}tH$
production, is the requirement of a more complicated trigger. The latter
provide hard isolated leptons as effective trigger signatures while the WBF
signal possesses a dijet and missing $p_T$ signature only. The two jets of the
WBF signal are very hard and widely separated, however, which allows for the
setup of a specific WBF trigger, with modest reliance on the Higgs decay
signature. The QCD dijet cross section for events with two jets of $p_T>40~
(20)$~GeV within $|\eta_j|<5$, separated by $ |\eta_{j_1} - \eta_{j_2} | >
4.4$, is about 2.3~(41)~$\mu$b. Requiring the presence of $\sla p_T$ with more
than 3 to 4$\sigma$ significance, a condition which is easily fulfilled by the
signal events with their $\sla p_T>100$~GeV, will immediately lower these
cross sections to the low nb level, {\em i.e.} to event rates of a few Hz only
at a luminosity of $10^{33}$cm$^{-2}$sec$^{-1}$. Similarly, the presence of
isolated electrons, muons or photons, of modest $p_T$, will allow for
efficient triggers of $H\to \tau \tau $, $H \to WW$ and $H\to \gamma\gamma$
events in WBF~\cite{dd:aa,dd:ww,dd:tt}. We strongly urge the LHC
detector collaborations to develop such triggers for a broad-based study of
WBF processes.


\acknowledgements
D.~Z.\ would like to thank the CERN theory group for its hospitality 
during the early stages of this work. Enlightening discussions with 
S.~Dasu, A.~Nikitenko, and D.~Rainwater are gratefully acknowledged. This
work was supported in part by the University of Wisconsin Research Committee
with funds granted by the Wisconsin Alumni Research Foundation, in part by the
U.~S.~Department of Energy under Contract No.~DE-FG02-95ER40896, by Conselho
Nacional de Desenvolvimento Cient\'{\i}fico e Tecnol\'ogico (CNPq), by
Funda\c{c}\~ao de Amparo \`a Pesquisa do Estado de S\~ao Paulo (FAPESP), and
by Programa de Apoio a N\'ucleos de Excel\^encia (PRONEX).


\bibliographystyle{plain}


\newpage


\begin{table}
\begin{tabular}{||c|c|c|c||}
 & signal $Hjj$  & QCD $Zjj$ and $Wjj$  & EW $Zjj$ and $Wjj$
\\
\hline
\hline
$P_{surv}$ & 0.87   & 0.28 & 0.82 
\end{tabular}
\medskip
\caption{Survival probabilities for the signal and background for a 
  veto of central jets with $p_T > 20$ GeV. From Ref.~\protect\cite{rainth}.
}
\label{surv}
\end{table}


\begin{table}
\begin{tabular}{||c|c|c|c|c|c||}
      	& QCD $Zjj$	& QCD $Wjj$	& EW $Zjj$ 	& EW $Wjj$ &total
\\
\hline
$\sigma$
	 	& 1254		& 1284		& 151		& 101	& 2790
\\
\hline
$P_{surv}\;\sigma$ 
		& 351		& 360		& 124		& 83	& 918

\\
\hline
$P_{surv}\;\sigma(\phi_{jj}<1)$ 
		& 71.8	& 70.2		& 14.8		& 9.9	& 167
\end{tabular}
\medskip
\caption{
Total cross sections (in fb) for the backgrounds after applying the cuts
(\protect\ref{cuts1}-\protect\ref{cuts2}) (first two lines) 
and (\protect\ref{cuts3}). 
In the last two lines we also include the central jet veto survival 
probabilities of Table~\protect\ref{surv}.
}
\label{ev:bckg}
\end{table}


\begin{table}
\begin{tabular}{||c|c|c|c|c|c|c|c||}
$M_H$ (GeV)   	& 110	& 120	& 130 	& 150	& 200 	& 300	& 400	
\\
\hline
$\sigma$ 
		& 282.  & 274.  & 266.  & 251.	& 214.	& 154.	& 110.
\\
\hline
$P_{surv}\;\sigma$ 
		& 245.  & 238.  & 232.  & 218.	& 186.	& 134.	& 95.7
\\
\hline
$P_{surv}\;\sigma(\phi_{jj}<1)$ 
		& 99.4	& 96.7	& 94.3	& 89.2	& 77.0  & 56.3	& 40.7
\end{tabular}
\medskip
\caption{
Same as Table \protect\ref{ev:bckg} for the signal at several invisible Higgs
masses, assuming Br($H\to \hbox{ invisible}) = 1$. Cross sections are 
given in fb.
}
\label{ev:sig}
\end{table}


\begin{table}
\begin{tabular}{||c|c|c|c|c|c|c|c||}
$M_H$ (GeV)   & 110	& 120	 & 130 	  & 150	   & 200   & 300  & 400	
\\
\hline
10 fb$^{-1}$  & 12.6\%  & 13.0\% & 13.3\% & 14.1\% & 16.3\%& 22.3\% & 30.8\%
\\
\hline
100 fb$^{-1}$ &  4.8\%  & 4.9\%  & 5.1\%  & 5.3\%  & 6.2\%& 8.5\% & 11.7\%
\end{tabular}
\medskip
\caption{
   Invisible branching ratio that can be probed at 95\% CL as a function 
   of $M_H$ for an integrated luminosity of 10 fb$^{-1}$ and 100 fb$^{-1}$. 
   A SM production cross section is assumed.  }
\label{res}
\end{table}


\newpage


\end{document}